# MORPHOLOGICAL AND STRUCTURAL CHARACTERIZATION OF $CrO_2/Cr_2O_3$ FILMS GROWN BY LASER-CVD

P.M. Sousa[1], A.J. Silvestre[2], N. Popovici[1] and O. Conde[1*]

[1]Departamento de Física, Universidade de Lisboa, Ed. C8, Campo Grande, 1749-016 Lisboa, Portugal

[2]Instituto Superior de Engenharia de Lisboa, R. Conselheiro Emídio Navarra, 1950-062 Lisboa, Portugal.

**Abstract**

This work reports on the synthesis of chromium (III, IV) oxides films by KrF laser-assisted CVD. Films were deposited onto sapphire substrates at room temperature by photodissociation of $Cr(CO)_6$ in dynamic atmospheres containing oxygen and argon. A study of the processing parameters has shown that partial pressure ratio of $O_2$ to $Cr(CO)_6$ and laser fluence are the prominent parameters that have to be accurately controlled in order to co-deposit both crystalline oxide phases. Films consistent with such a two-phase system were synthesised for a laser fluence of 75 mJ cm$^{-2}$ and a partial pressure ratio about 1.

---

[*] Corresponding author: Fax.:+(351) 217 500 977, E-mail: oconde@fc.ul.pt

# 1. Introduction

Chromium oxide thin films are of great interest due to their wide variety of technological applications. The most stable phase is the insulating antiferromagnetic $Cr_2O_3$. This oxide has several applications, for instance in catalysis [1] and solar thermal energy collectors [2]. Currently, low-reflective $Cr_2O_3$/Cr films are widely used as black matrix films in liquid crystal displays [2]. $Cr_2O_3$ films are also used as protective films because of their good chemical and wear resistance, and low friction coefficient [3]. In contrast, chromium dioxide ($CrO_2$) is strongly ferromagnetic at room temperature ($T_C = 393$ K) and has a half-metallic band structure fully spin-polarised at the Fermi level [4-7], making it extremely attractive for use in magnetoelectronic devices. Furthermore, $Cr_2O_3$/$CrO_2$ biphasic films show remarkable magnetoresistance due to intergrain tunnelling between high spin polarized crystals [8,9]. However, the synthesis of chromium (III, IV) oxides mixtures at temperatures compatible with semiconductor processing has been a difficult task, due to the metastability of $CrO_2$ which easily decomposes into the insulating antiferromagnetic $Cr_2O_3$ phase. Therefore, there is a continuous search for deposition methods that allow the growth of these films at low temperatures.

Laser-assisted CVD is a selective area deposition technique which has the potential to meet the requirements stated above. In a few previous studies, Dowben and co-authors [10-13] carried out the photolytic decomposition of $Cr(CO)_6$ by nitrogen laser radiation ($\lambda = 337$ nm), in static conditions, having shown the synthesis of films containing both $Cr_2O_3$ and $CrO_2$ phases. Looking for a better efficiency of the deposition process, it seems to be advantageous to explore dynamic atmospheres and to use other UV wavelengths for which $Cr(CO)_6$ absorption cross section is much higher than for the nitrogen laser, *e.g.* KrF laser radiation ($\lambda = 248$ nm) [14,15].

In this paper, we present results on the synthesis of chromium (III, IV) oxides films produced by KrF laser-assisted CVD. Films were grown onto $Al_2O_3$ (0001) substrates at room temperature by photodissociation of $Cr(CO)_6$ in dynamic atmospheres containing $O_2$ and Ar.

**2. Experimental details**

The experimental apparatus used for the experiments consists of a KrF excimer laser, a stainless steel HV chamber, a HV pumping system and a gas system for delivering the reactants to the reaction chamber. The KrF laser ($\lambda = 248$ nm, $\tau = 30$ ns) has a beam dimension of 24×12 mm$^2$. The laser beam impinges on the substrate at perpendicular incidence and by using an adequate optical system, it is possible to control the beam spot size and thus the fluence at the substrate surface. The laser fluences reported in table I have been evaluated taking into account both the beam energy loss on the optical components and the energy loss due to absorption by the reactant gas mixture. Prior to any deposition experiment, the reactor was always evacuated to a base pressure lower than $5 \times 10^{-4}$ Pa. All the deposition experiments were conducted in dynamic regime. Ar was used as buffer gas and to flash the inner face of the chamber window used for transmission of the laser beam.

Chromium hexacarbonyl powder was sublimated in a controlled temperature stainless steel cell connected to the LCVD reactor. The $Cr(CO)_6$ vapour pressure is given by the following equation [9]: $\log_{10}P[Pa] = 12.75 - 3285/T[K]$. Oxygen was used as carrier gas, except in the experiments carried out in a non-oxidant atmosphere where it was replaced by argon. The partial pressures of $Cr(CO)_6$, $O_2$ and Ar inside the reactor are settled by the cell temperature, the flux of the carrier gas, the flux of Ar and the total working pressure, assuming a viscous flow regime between the cell and the reactor. The chromium oxide films were grown at room temperature on $Al_2O_3$ (0001) substrates for a laser pulse repetition rate of 5 Hz and total pressure $P_T = 50$ Pa. The overall experimental parameters are summarised in table I.

In order to study the chromium oxide phases synthesised as a function of laser fluence and partial pressure ratio of $O_2$ and $Cr(CO)_6$, structural analysis of the as-deposited films was carried out by X-ray diffraction with Cu K$\alpha$ radiation and by micro-Raman spectroscopy using the 514.5 nm excitation line of an $Ar^+$ laser. The morphology and microstructure of the films were analysed by optical microscopy (OM) and scanning electron microscopy (SEM), and their thickness measured by stylus profilometry.

## 3. Results and discussion

### 3.1. Surface morphology and microstructure

The as-deposited films present good adherence and are approximately rectangular in shape following the laser beam profile. OM observations revealed that they are quite homogeneous and exhibit a colour varying between greenish and nearly black, depending on the processing parameters used on their synthesis. Since $Cr_2O_3$ has a characteristic green colour while $CrO_2$ is black, film colour can be used as first control to infer the presence of chromium dioxide in the deposited material. Films produced with partial pressure ratio $p_{O2}/p_{Cr} = 1$ tend to be darker, otherwise the greenish colouration tends to be dominant. Also the microstructure of the films may be related with both chromium oxide phases. As documented in literature [16], films of $Cr_2O_3$ often consist of spheroid particles, whereas $CrO_2$ usually appears as rod-shaped particles. SEM analysis of our samples shows a microstructure consisting mainly of granular and randomly oriented rod-shaped particles, the latter prevailing in the darker films synthesised with $p_{O2}/p_{Cr} = 1$ (Fig. 1). As it will be discussed hereafter, the structural analysis of the samples strongly supports that films produced with $p_{O2}/p_{Cr} = 1$ are those with higher content of $CrO_2$.

### 3.2. Structural analysis

#### 3.2.1. Effect of gas phase composition

Figure 2 shows X-ray diffraction patterns of films deposited during 4 hours with F = 75 mJ cm$^{-2}$ and $p_{O2}/p_{Cr}$ ratios ranging from 0 to 4.4. For films grown at $p_{O2}/p_{Cr} < 1$ (Fig. 2a and 2b), only the $Cr_2O_3$ crystallographic planes match the diffractograms and no traces of crystalline $CrO_2$ phase were observed. Also the film deposited with $p_{O2}/p_{Cr} > 1$ (Fig. 2d) shows no evidence of chromium dioxide deposition. The formation of both polycrystalline $Cr_2O_3$ and $CrO_2$ phases was achieved in samples grown with $p_{O2}/p_{Cr} = 1$ (Fig. 2c). In this diffractogram, both the characteristic diffraction peaks of $Cr_2O_3$ and the (110) and (111) diffraction lines of $CrO_2$ can be clearly seen. Although the role of oxygen on the chemical reaction that leads to the co-deposition of both chromium oxide phases by photodecomposition of $Cr(CO)_6$ is not completely understood, our results show that oxygen is absolutely necessary to synthesise $CrO_2$ but not to achieve $Cr_2O_3$ deposition. In fact, the latter phase can be grown without oxygen (Fig.2a) as a result of the recombination of Cr with the free oxygen released from the decomposition of $Cr(CO)_6$ at low temperature [17,18]. Dowben and co-workers [11,12] have pointed out that the partial pressure of oxygen is the prominent parameter to be controlled in order to tailor the relative amount of $Cr_2O_3$ and $CrO_2$ that result from the dissociation of $Cr(CO)_6$, suggesting that an increase of the $p_{O2}/p_{Cr}$ ratio should allow to improve the formation of $CrO_2$. Our results show that co-deposition of both polycrystalline oxide phases is only obtained when the precursors partial pressure ratio is close to unity. On the other hand, the observed broadening of the $Cr_2O_3$ diffraction peaks when $p_{O2}/p_{Cr}$ increases suggests that higher $p_{O2}/p_{Cr}$ ratios induce some degree of amorphisation of this phase, similarly to what has been observed in CVD processes of $Cr_2O_3$ using the same precursors as in the present work [19].

### *3.2.2. Effect of laser fluence*

Besides partial pressures ratio, also laser fluence plays an important role on the phase composition of the deposited material. When absorbing two photons at 248 nm, an isolated $Cr(CO)_6$ molecule should decompose to Cr and 6 CO with a quantum yield given in literature as ~1 [20]. By reacting with $O_2$, the photodissociated Cr species would give rise to $CrO_2$. However, the mechanism by which the photodissociation of $Cr(CO)_6$ may lead to chromium dioxide is quite complex and still not clear. A simplified scheme can be drawn as follows: a single 248 nm photon leads to the removal of two carbonyl ligands resulting in $Cr(CO)_4$ formation [14,20,21]. At laser fluence higher than 5 mJ cm$^{-2}$ a second photon gives rise to secondary photodissociation of Cr(CO)$_4$ into Cr* and 4CO. Then, the formation of CrO$_2$ proceeds via the reaction of the excited chromium with oxygen. Surface reactions at the substrate surface such as 2CrO$_2$ → Cr$_2$O$_3$ + 1/2O$_2$ may occur, the final composition of the deposited material depending on the extension of this reaction. Nevertheless, these features are only true for a very low pressure of Cr(CO)$_6$, for which collisions are absent. In systems where collisions play an important role, the statistical fragmentation processes that follow the initial photodissociation event may be inhibited by collision vibrational relaxation of the photoproducts. As we aim at achieving co-deposition of both Cr$_2$O$_3$ and CrO$_2$ at a significant growth rate, higher pressures are required and, consequently, higher laser fluences should be used. For the range of pressures given in this work, chromium oxide deposition was obtained only for laser fluences higher than 50 mJ cm$^{-2}$.

Figure 3 shows X-ray diffraction patterns (Fig. 3a) and micro-Raman spectra (Fig. 3b) of films deposited during 4 hours with $p_{O2} / p_{Cr} = 1$ and various laser fluences. Each Raman spectrum was normalised to its Cr$_2$O$_3$ A$_{1g}$ peak intensity to ease the comparison among the various spectra. As previously described, the XRD pattern of the film grown with F = 75 mJ cm$^{-2}$ (Fig. 3a, film I) clearly matches the (110) and (111) diffraction planes of CrO$_2$ as well as the diffraction lines of Cr$_2$O$_3$. Confirming the synthesis of chromium (III, IV) oxides, the micro-Raman spectrum recorded over the same sample (Fig. 2b, film I) reveals an intense peak assigned to the B$_{2g}$ mode of CrO$_2$, at 701.5 cm$^{-1}$ Raman shift, in addition to the symmetric bands of Cr$_2$O$_3$ [22,23]. By decreasing the laser fluence, film thickness decreases and a higher degree of amorphisation is induced in the deposited material. In fact, on the XRD pattern of the film synthesised with F = 60 mJ cm$^{-2}$ (Fig. 3a, film II) only the (104) and (110) diffraction peaks of Cr$_2$O$_3$ can be clearly identified, and these peaks are broader and less intense than for film I. Concerning the chromium dioxide, the XRD data indicate that either it was not formed at this fluence or it formed with a level of amorphisation extremely high. However, the micro-Raman spectrum of the same film (Fig. 3b, film II) provides evidence that co-deposition of $CrO_2$ and $Cr_2O_3$ still occurs at F = 60 mJ cm$^{-2}$: besides the $Cr_2O_3$ peaks, also the $CrO_2$ $B_{2g}$ band is present in the Raman spectrum. Table II summarises the principal features of the $Cr_2O_3$ $A_{1g}$ and $CrO_2$ $B_{2g}$ Raman bands evaluated by fitting the spectra of Fig. 3b with Lorentzian curves. By comparing the data for films I and II, the amorphisation of both oxides is well observed via the increase of the full-with at half-maximum (FWHM) of the $Cr_2O_3$ $A_{1g}$ and $CrO_2$ $B_{2g}$ symmetric modes for film II. Besides amorphisation, a low laser fluence also yields a low relative content of $CrO_2$ since the intensity ratio I($CrO_2$ $B_{2g}$)/I($Cr_2O_3$ $A_{1g}$) for film II is about 12% lower than for film I.

On the other hand, the increase of laser fluence to 100 mJ cm$^{-2}$ favours the deposition of crystalline $Cr_2O_3$. The FWHM of the band at ~555 cm$^{-1}$ decreases from film I to film III and the $Cr_2O_3$ $E_g$ bands are better resolved for film III (Fig. 3b). The absence of $CrO_2$ diffraction lines on pattern III of figure 3a may be due to an extended transformation of the initially formed $CrO_2$ into $Cr_2O_3$ caused by heating of the film/substrate during the deposition process at higher laser fluences. Indeed, films processed with F > 120 mJ cm$^{-2}$ showed evidence either of surface melting or phase transformation of the sapphire substrate.

From XRD and micro-Raman structural analysis, one can conclude that narrow windows for the laser fluence and $p_{O2}/p_{Cr}$ ratio, centred respectively at 75 mJ cm$^{-2}$ and 1, have to be used in order to grow crystalline chromium oxide thin films containing both $CrO_2$ and $Cr_2O_3$ phases.

### 3.3. Growth rate

Films prepared with the experimental conditions mentioned above present near flat thickness profiles and their thickness increases linearly with deposition time (Fig. 4), giving an apparent deposition rate of 3.2 nm min$^{-1}$. The intersection of the straight line with the time axis gives a characteristic time $\Delta t \cong 50$ min, which is very close to the time delay observed experimentally between the instant the laser is turned on and the abrupt change of the sample reflectivity.

## 4. Conclusion

Films displaying a behaviour consistent with a two-phase system of $Cr_2O_3$ and $CrO_2$ oxides were produced by KrF laser photodissociation of $Cr(CO)_6$ in dynamic atmospheres containing oxygen and argon. This technique seems to open an interesting approach to the synthesis of biphasic chromium (III, IV) oxides films at room temperature. It was shown that partial pressures ratio of the precursors and laser fluence are the prominent parameters that have to be accurately controlled in order to co-deposit both crystalline oxide phases. Biphasic films with both $CrO_2$ and $Cr_2O_3$ phases were grown for laser fluence and $p_{O2}/p_{Cr}$ ratio values centred at 75 mJ cm$^{-2}$ and 1, respectively. For these films a deposition rate of 3.2 nm min$^{-1}$ was measured.

**Acknowledgements**

This work was supported by the EU contract FENIKS: G5RD-CT-2001-00535

TABLE I

LCVD experimental conditions used to deposit chromium oxide films

| Experimental parameters | | Estimated parameters | |
|---|---|---|---|
| Beam energy, E [mJ] | 4.1 – 15.6 | Fluence at the substrate, F [mJ cm$^{-2}$] | 50 – 190 |
| Spot area, S [mm$^2$] | 8.2 | $Cr(CO)_6$ flux, $\Phi_{Cr}$ [sccm] | 1.1 – 1.3 |
| $Cr(CO)_6$ cell temperature, $T_{cell}$ [ºC] | 19 – 22 | $Cr(CO)_6$ partial pressure, $p_{Cr}$ [Pa] | 0.7 – 2.3 |
| $O_2$ carrier flux, $\Phi_{O2}$ [sccm] | 1 – 10 | $O_2$ partial pressure, $p_{O2}$ [Pa] | 0.0 – 8.0 |
| Ar flux, $\Phi_{Ar}$ [sccm] | 50 | Ar partial pressure, $p_{Ar}$ [Pa] | 40.1 – 48.8 |
| Deposition time, $t_{dep}$ [min] | 240 – 360 | | |

TABLE II

Main features of the $Cr_2O_3$ $A_{1g}$ and $CrO_2$ $B_{2g}$ Raman bands of films referred to Fig. 3b.

| Film reference | Band position (cm$^{-1}$) | | FWHM (cm$^{-1}$) | | Intensity ratio |
|---|---|---|---|---|---|
| | $Cr_2O_3$ ($A_{1g}$) | $CrO_2$ ($B_{2g}$) | $Cr_2O_3$ ($A_{1g}$) | $CrO_2$ ($B_{2g}$) | $CrO_2$ ($B_{2g}$)/$Cr_2O_3$ ($A_{1g}$) [a] |
| Film I   | 554.9 | 701.5 | 11.0 | 29.5 | 0.542 |
| Film II  | 551.5 | 699.5 | 24.6 | 35.7 | 0.478 |
| Film III | 555.1 | 700.6 | 10.4 | 29.6 | 0.524 |

**Figure captions**

Figure 1 – SEM image of a film deposited during 4 hours with the following deposition parameters: $P_T$ = 50 Pa, $\Phi_{Cr}$ = 1 sccm, $\Phi_{O2}$ = 1 sccm, $\Phi_{Ar}$ = 50 sccm and F = 75 mJ cm$^{-2}$.

Figure 2 – XRD patterns of films deposited during 4 hours at total pressure of 50 Pa, F = 75 mJ cm$^{-2}$ and various $p_{O2}/p_{Cr}$ ratios: a) $p_{O2}/p_{Cr}$ = 0; b) $p_{O2}/p_{Cr}$ = 0.7; c) $p_{O2}/p_{Cr}$ = 1; d) $p_{O2}/p_{Cr}$ = 4.4.

Figure 3 – a) XRD patterns and b) micro-Raman spectra of films deposited during 4 hours with different fluences. Deposition parameters: $P_T$ = 50 Pa, $\Phi_{Cr}$ = 1 sccm, $\Phi_{O2}$ = 1 sccm and $\Phi_{Ar}$ = 50 sccm. Film I, F = 75 mJ cm$^{-2}$; film II, F = 60 mJ cm$^{-2}$; film III, F = 100 mJ cm$^{-2}$. Each Raman spectrum was normalised to its $Cr_2O_3$ $A_{1g}$ peak intensity.

Figure 4 – Film thickness *vs.* deposition time for chromium (III, IV) oxides films processed with the following parameters: $P_T$ = 50 Pa, $\Phi_{Cr}$ = 1 sccm, $\Phi_{O2}$ = 1 sccm, $\Phi_{Ar}$ = 50 sccm and F = 75 mJ cm$^{-2}$. ■, experimental data; -----, curve fit.

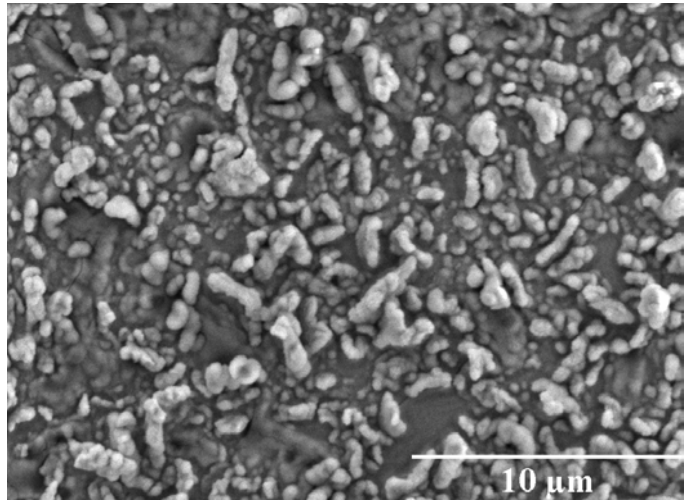

Figure 1

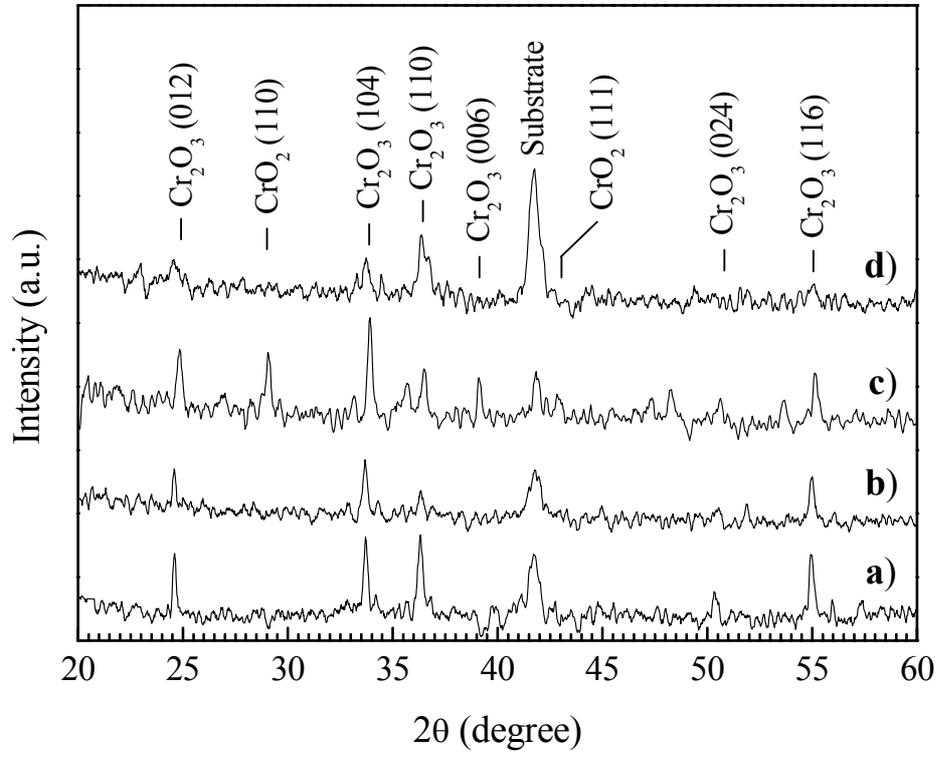

Figure 2

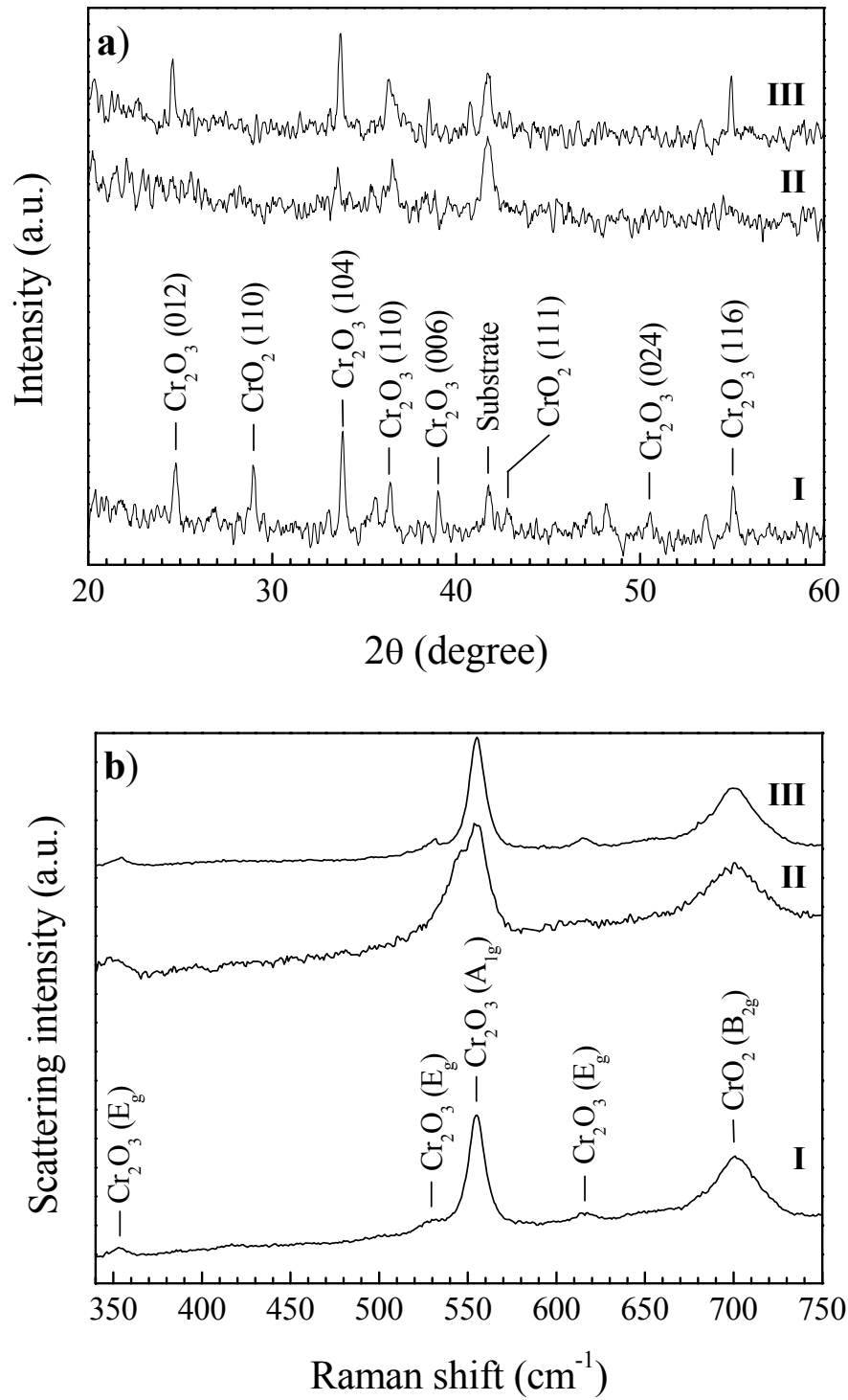

Figure 3

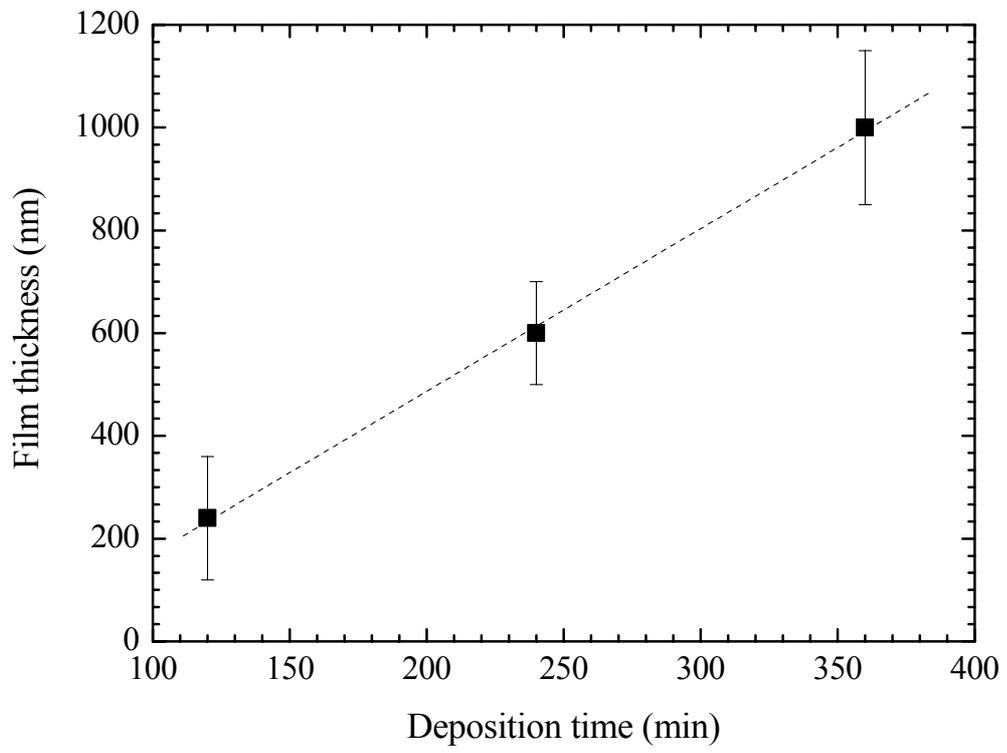

Figure 4